\documentclass[manuscript=letter]{achemso}

\setkeys{acs}{articletitle = true}

\usepackage{graphicx}
\usepackage{dcolumn}
\usepackage{cancel}
\usepackage{ulem}
\setcitestyle{square}

\usepackage{bm}
\usepackage{verbatim}

\usepackage{bm,nicefrac}

\usepackage{mathtools}

\usepackage{amsmath}

\usepackage{subfig}

\newsavebox{\measurebox}

\usepackage{braket}

\usepackage{asymptote}

\usepackage{siunitx}

\usepackage[utf8]{inputenc}

\usepackage{tabu,multirow}
\usepackage{makecell}
\usepackage{boldline}

\usepackage{chemfig}

\title{Non-Equilibrium Thermodynamics of Charge Separation in Organic Solar Cells}

\author{W.~Kaiser}
 \affiliation{Department of Electrical and Computer Engineering, Technical University of Munich, Karlstra{\ss}e 45, 80333 Munich, Germany}
\author{V.~Jankovi\'c}
 \affiliation{Scientific Computing Laboratory, Center for the Study of Complex Systems, Institute of Physics Belgrade, University of Belgrade, Pregrevica 118, Belgrade 11080, Serbia}
\author{N.~Vukmirovi\'c}
 \affiliation{Scientific Computing Laboratory, Center for the Study of Complex Systems, Institute of Physics Belgrade, University of Belgrade, Pregrevica 118, Belgrade 11080, Serbia}
\author{A.~Gagliardi}
\email{alessio.gagliardi@tum.de}
 \affiliation{Department of Electrical and Computer Engineering, Technical University of Munich, Karlstra{\ss}e 45, 80333 Munich, Germany}


\begin{document}

\begin{abstract}
This work presents a novel theoretical description of the non-equilibrium thermodynamics of charge separation process in organic solar cells (OSCs). Using the theory of stochastic thermodynamics, we connect the phonon-assisted dynamics and recombination of electron--hole pairs within a photo-excited organic bilayer with the thermodynamic free energy. We analyze the impact of energetic disorder and delocalization on the free energy, average energy and entropy. For high energetic disorder, the site population is well described by equilibrium. We observe significant deviations from equilibrium for delocalized electron--hole pairs at small energetic disorder, representing efficient OSCs. Our results emphasize that both a large Gibbs entropy and large initial separation are required to achieve efficient charge separation. A decrease in free energy barrier with increased distances between charges does not necessarily correlate with the separation yield. Our presented framework can further shed light on the transient thermodynamic free energy, allowing previously inaccessible insight into the individual thermodynamic contributions from energy and entropy on sub-ns timescales. Transient simulations reveal large Gibbs entropy on ps-timescales for even highest disorder, which may explain the efficient separation of "hot" CT states. 
\end{abstract}

\maketitle

\section{Introduction}
\label{sec:intro}

Despite years of research on the fundamental properties of organic solar cells (OSCs), one of their essential aspects is still not fully understood: what causes the efficient dissociation of charge transfer (CT) states into free charge carriers despite the presence of the strong Coulomb attraction? Various mechanisms have been proposed to promote CT separation. Delocalized charge carriers can efficiently escape their strong binding and enhance the separation of CT states due to a reduced Coulomb interaction \cite{kahle2018does,deibel2009origin,tscheuschner2015combined,jankovic2018combination,athanasopoulos2019disorder,felekidis2020role}. "Hot" CT states are considered to provide a sufficient initial excess energy to overcome the binding energy \cite{bassler2015hot,grancini2013hot,jones2014hot,bakulin2012role}. Previous studies emphasize the role of the thermodynamic free energy barrier of CT states during charge separation under the assumption of equilibrated site occupations.\cite{gregg2011entropy, hood2016entropy,PhysRevLett.114.128701}

In many organic semiconductors, the energy landscape is strongly disordered due to the lack of long-range order \cite{kohler2015electronic}. Consequently, excitations in form of electronic charges or singlet excitons are usually initially out of equilibrium and take a certain time to relax within the density of states (DOS). Consequently -- especially on fast timescales -- the relaxation might be incomplete, implying non-equilibrium charge \cite{nikitenko2007non, melianas2015photo, melianas2017non} and exciton \cite{ansari2018theoretical, athanasopoulos2009exciton} dynamics which equilibrium theories cannot accurately describe. Recent studies discuss the nature of CT states at interfaces in OSCs. Non-equilibrated electron--hole pairs due to incomplete thermalization were observed in CT electroluminescence \cite{melianas2019nonequilibrium} and photoluminescence measurements \cite{brigeman2018nonthermal} in bulk heterojunction solar cells, while Neher and coworkers reported that equilibrated charge carriers govern nongeminate recombination in steady-state \cite{roland2019equilibrated}. In this concern, understanding the role of non-equilibrium CT state dynamics is crucial to obtain a reliable interpretation of the thermodynamic free energy barrier of photogenerated electron--hole pairs during charge separation.

Gregg studied the relation between the dimensionality and entropy based on the degeneracy of states $W$ as a function of the separation distance $r$ using the equilibrium relation $\Delta S(r) = k_B \ln W(r)$ \cite{gregg2011entropy}. In materials of higher dimensionality, the free energy barrier is significantly reduced due to the increase in available states.~\cite{gregg2011entropy,ChemRev.110.6736,JPhysChemLett.4.1707} Hood and Kassal emphasized that relying on dimensionality captures the case of semiconductors without energetic disorder, while it underestimates the contribution of the energetic disorder to the free energy for disordered organic semiconductors \cite{hood2016entropy}. To account for the DOS, they calculated the free energy based on the canonical partition function assuming equilibrated CT states. The partition function is purely based on the difference between the initial and final energies of the CT states and the degeneracy of final configurations after separation. This equilibrium method predicts a reduction in free energy barrier with increased disorder $\sigma$ and proposes spontaneous charge separation for $\sigma \geq \SI{100}{\milli\electronvolt}$. However, it is known that disorder can hamper charge transport and trap charge carriers. Additionally, an increased disorder impacts recombination of CT states \cite{albes2017influence}. Using a fully quantum and statistical approach, Lankevich and Bittner studied the combined influence of Coulomb attraction, disorder, carrier delocalization, and carrier--phonon couplings on charge separation thermodynamics.~\cite{lankevich2018Voc} Their work links the "hot" and "cold" separation mechanisms, which are respectively obtained under the assumption of microcanonical and canonical distribution of CT populations in energy. 

Some authors claim that charge separation is a strongly non-equilibrium process and that it requires a non-equilibrium description \cite{shi2017enhancement,giazitzidis2014charge}. Giazitzidis \textit{et al.} developed a simplified non-equilibrium description by correcting the equilibrium free energy with the probability distribution of finding a CT state at a certain distance obtained from Monte Carlo simulations \cite{giazitzidis2014charge}. Shi \textit{et al.} emphasize that the disorder-enhanced dissociation is a non-equilibrium effect \cite{shi2017enhancement}. They showed a deviation between the average energy barrier obtained from the equilibrium description and the non-equilibrium energy barrier by averaging over Monte Carlo trajectories, but neglected the weighting of the energies by their population probability. 

Existing studies do not fully capture the non-equilibrium nature of charge separation on the thermodynamic free energy and consequently do not allow insight in whether and to which extent equilibrium methods provide a valid description. In this work, we study the non-equilibrium thermodynamics of CT separation in a one-dimensional model Hamiltonian of a photo-excited organic bilayer, representing the donor--acceptor interface in OSCs, based on the theory of stochastic thermodynamics~\cite{van2015ensemble,seifert2012stochastic}. The non-equilibrium site occupation is obtained from the stationary state that reflects the interplay between exciton generation by natural sunlight, phonon-assisted exciton and charge transport, recombination, and extraction of fully separated charges. The non-equilibrium free energy is derived based on the Kullback--Leibler divergence between the non-equilibrium and the equilibrium probability distribution in energy and position of CT states. The presented method is finally used to study the impact of energetic disorder and delocalization on free energy. 

\section{Method}
\label{sec:method}

Our analysis is based on a one-dimensional model Hamiltonian of an organic bilayer. This model captures the charge separation dynamics under the impact of energetic disorder, charge delocalization, electric field, and the interaction between charge carriers and phonons \cite{jankovic2018combination}, see also Supporting Information for details of the model and the values of its parameters. Electron--hole pairs evolve within a phase space of available states with large energetic disorder and a variable mean distance between the wave function of the electron and the hole. We assume that charge separation starts from states of donor excitons that are populated as a result of the absorption of natural sunlight. It proceeds via a sequence of phonon-assisted transitions between different exciton states that finishes once the charges are collected or recombine. The electron and hole are considered collected once they are fully separated in space such that their wave functions are sufficiently localized in a defined contact region $l_c$ at the end of the acceptor and the donor, respectively. The stationary population $f_x^0$ of exciton state $|x\rangle$ is obtained by solving
\begin{equation}
\label{Eq:ss}
    0=g_x-\tau_x^{-1} f_x^0-\sum_{x'}w_{x'x} f_x^0+\sum_{x'\notin C}w_{xx'}f_{x'}^0
\end{equation}
where $C$ denotes states of fully separated charges from which the collection occurs (contact states), $w_{x'x}$ is the transition rate from $|x\rangle$ to $|x'\rangle$, while $\tau_x$ is the lifetime of state $|x\rangle$. The generation rate $g_x$ is assumed to be non-zero only for states of donor excitons and is given by the Weisskopf--Wigner formula~\cite{Scully-Zubairy-book}
\begin{equation}
\label{Eq:ww-gen-rate}
    g_x = \vert\mu_x\vert^2\frac{\left(\beta_\mathrm{R}E_x\right)^3}{\exp\left(\beta_\mathrm{R}E_x\right) - 1}\,,
\end{equation}
where $\mu_x$ is the matrix element for the optical excitation of state $\ket{x}$ from the ground state, $E_x$ is the energy of state $\ket{x}$, while $\beta_\mathrm{R}^{-1}=k_\mathrm{B}T_\mathrm{R}$ with $T_\mathrm{R}=\SI{5.8e3}{\kelvin}$ being the radiation temperature of the sun. The matrix element $\mu_x$ was calculated according to
\begin{equation}
    \mu_x = \sum_{i}\psi^{x*}_{i}d_i^{\mathrm{cv}}\,,
\end{equation}
where $\psi^{x}_{i}$ is the component of the excitonic wave functions at site $i$, while $d^\mathrm{cv}_i = 0$ in the acceptor region and $d^\mathrm{cv}_i = \text{const}$ in the donor part of the bilayer. The expression for the charge separation efficiency
\begin{equation}
\eta=\frac{\sum_{x'\in C}\sum_{x\notin C}w_{x'x}f_x^0}{\sum_{x\notin C}g_x}
\end{equation}
contains phonon-assisted transition rates $w_{x'x}$ toward state $|x'\rangle$ of fully separated charges, which are appreciable only when state $|x\rangle$ features the electron in the acceptor and the hole in the donor part of the bilayer. In the following, instead of considering the full distribution, we concentrate on the stationary distribution of CT states, in which the charges are spatially separated and reside in different regions of the bilayer.

In recent years, the theory of stochastic thermodynamics has been developed and successfully applied to small-scale systems, especially to study the effect of fluctuations \cite{van2015ensemble,seifert2012stochastic,rao2016nonequilibrium,leonard2013stochastic}. This novel theory provides a powerful description of the thermodynamics of stochastic processes governed by Markovian dynamics and can be applied to a broad class of non-equilibrium systems and processes. To study if, and to which extent, charge separation occurs out of equilibrium, we formulate the stochastic free energy $F_\mathrm{neq}$ based on the Kullback--Leibler divergence. \cite{van2015ensemble,qian2001relative}

To provide a comparable analysis to Hood and Kassal \cite{hood2016entropy}, we compute the non-equilibrium free energy $F_\mathrm{neq}(r)$ as a function of the intrapair distance $r$ using the Kullback--Leibler divergence of all CT states $m$ with intrapair distance $r_m\in\left[ r-b/2 , r+b/2\right)$:
\begin{equation}
    \beta \left( F_\mathrm{neq}(r) - F_\mathrm{eq}(r) \right)= \sum_{ \left\lbrace m | r_m \in [r - b/2, r + b/2) \right\rbrace} p^\mathrm{neq}_{m} \ln\left(\frac{p^\mathrm{neq}_{m}}{p^\mathrm{eq}_{m}}\right)\,.\label{eq:f_neq}
\end{equation}
For a given $r$, the non-equilibrium distribution $p_m^\mathrm{neq}$ of CT states whose intrapair separation $r_m\in[r-b/2,r+b/2)$ is obtained from the stationary populations $f_m^0$, entering Eq.~\ref{Eq:ss}, as follows
\begin{equation}
\label{Eq:neq-dist}
   p_m^\mathrm{neq}=f_m^0\left(\sum_{ \left\lbrace m | r_m \in [r - b/2, r + b/2) \right\rbrace}f_m^0\right)^{-1} \,.
\end{equation}
At the same time, the equilibrium distribution $p_m^\mathrm{eq}$ is obtained using the approach of Hood and Kassal~\cite{hood2016entropy}
\begin{equation}
    p_m^\mathrm{eq}=\exp(-\beta E_m)\left[\sum_{ \left\lbrace m | r_m \in [r - b/2, r + b/2) \right\rbrace}\exp{\left(-\beta E_m\right)}\right]^{-1}
\end{equation}
where $\beta^{-1}=k_\mathrm{B} T$ with $T$ being the temperature of the phonon bath. In all the results to be presented, we take $b=\SI{5}{\nano\meter}$. Rearranging Eq.~\ref{eq:f_neq} gives the well-known expression of the free energy:
\begin{align}
    F_\mathrm{neq}(r) &= \sum_{m} p^\mathrm{neq}_{m} E_m + \beta^{-1} p^\mathrm{neq}_{m}\ln p^\mathrm{neq}_{m} = E_\mathrm{neq}(r) - TS_\mathrm{neq}(r)\,.\label{eq:final}
\end{align}
Using the non-equilibrium probability distribution $p^\mathrm{neq}$, we can calculate the average energy $E_\mathrm{neq}$, which CT states pass during the separation process, as well as the entropy contribution $TS_\mathrm{neq}$ to the free energy. $S_\mathrm{neq}$ has the form of the Gibbs entropy and accounts for the accessible states during charge separation. 

One major advantage of the presented model is that it can be extended to study the time dependence of the thermodynamic measures as the initially unexcited bilayer approaches the stationary state. The final-value theorem for the Laplace transform guarantees that the stationary solution $\boldsymbol{f}^0$ of Eq.~\ref{Eq:ss} can also be obtained by integrating the temporal counterpart of Eq.~\ref{Eq:ss},
\begin{equation}
\label{Eq:temporal-counterpart}
    \frac{d\boldsymbol{f}(t)}{dt}=\boldsymbol{g}-\widehat{W}\boldsymbol{f}(t)
\end{equation}
up to sufficiently long times. The generation vector $\boldsymbol{g}$ contains all donor exciton generation rates $g_x$ (Eq.~\ref{Eq:ww-gen-rate}), while the matrix $\widehat{W}$ contains all rates of recombination and phonon-assisted transport.
The solution of Eq.~\ref{Eq:temporal-counterpart}
\begin{equation}
\label{Eq:temporal-solution}
    \boldsymbol{f}(t)=\int_0^t d\tau\:e^{-\widehat{W}(t-\tau)}\boldsymbol{g}
\end{equation}
with the initial condition $\boldsymbol{f}(t=0)=\boldsymbol{0}$, can be seen as a superposition of all possible evolutions that are triggered by sudden excitations at instants $\tau\in[0,t]$ and subsequently followed during the time interval of length $t-\tau$. The individual evolutions are representative of the dynamics observed in time-resolved experiments, in which donor exciton states are suddenly excited according to the generation vector $\boldsymbol{g}$.~\cite{JChemPhys.153.244110} Correspondingly, using $\boldsymbol{f}(t)$ to compute the time-dependent nonequilibrium distribution $p_m^\mathrm{neq}(t)$ (Eq.~\ref{Eq:neq-dist}) allows us to examine the transient behavior of thermodynamic quantities on intrinsic time scales of the charge separation process. To compute $\boldsymbol{f}(t)$, we use an efficient numerical scheme that has recently been developed in Ref~\citenum{jankovic2020energy} and that is briefly summarized in the Supporting Information.

\section{Results and Discussion}
\label{sec:results}

\begin{figure}[t!]
\centering
\sbox{\measurebox}{%
  \begin{minipage}[b]{.45\textwidth}
  \subfloat{\label{fig:figA}\includegraphics[width=\textwidth]{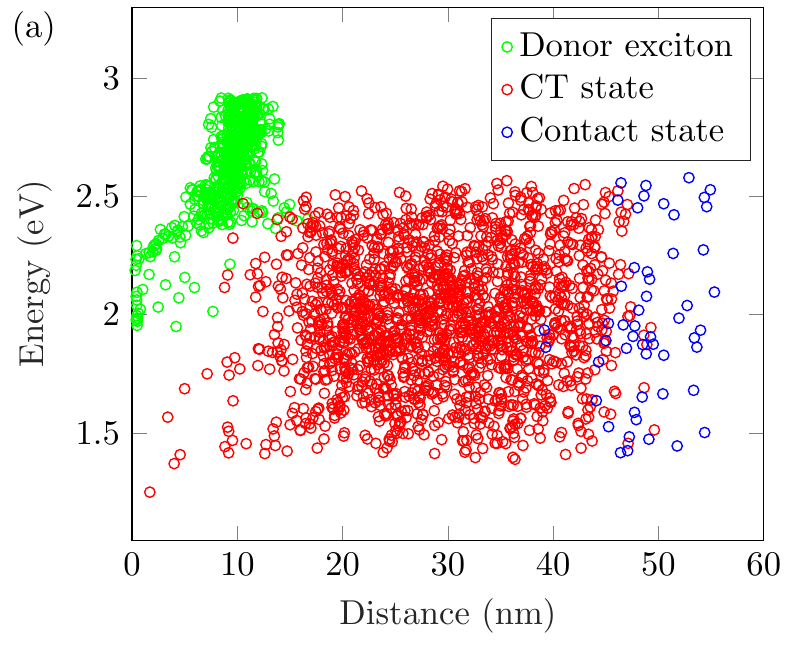}}
    \vfill
    \subfloat{\label{fig:figC}\includegraphics[width=\textwidth]{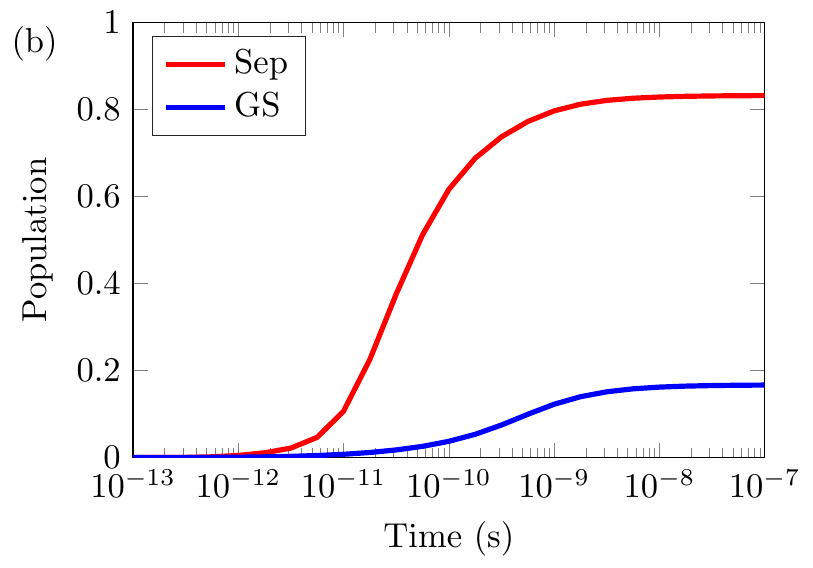}}
  \end{minipage}}
\usebox{\measurebox}
\qquad
\begin{minipage}[b][\ht\measurebox][s]{.42\textwidth}
\centering
\subfloat{\label{fig:figB}\includegraphics[width=\textwidth]{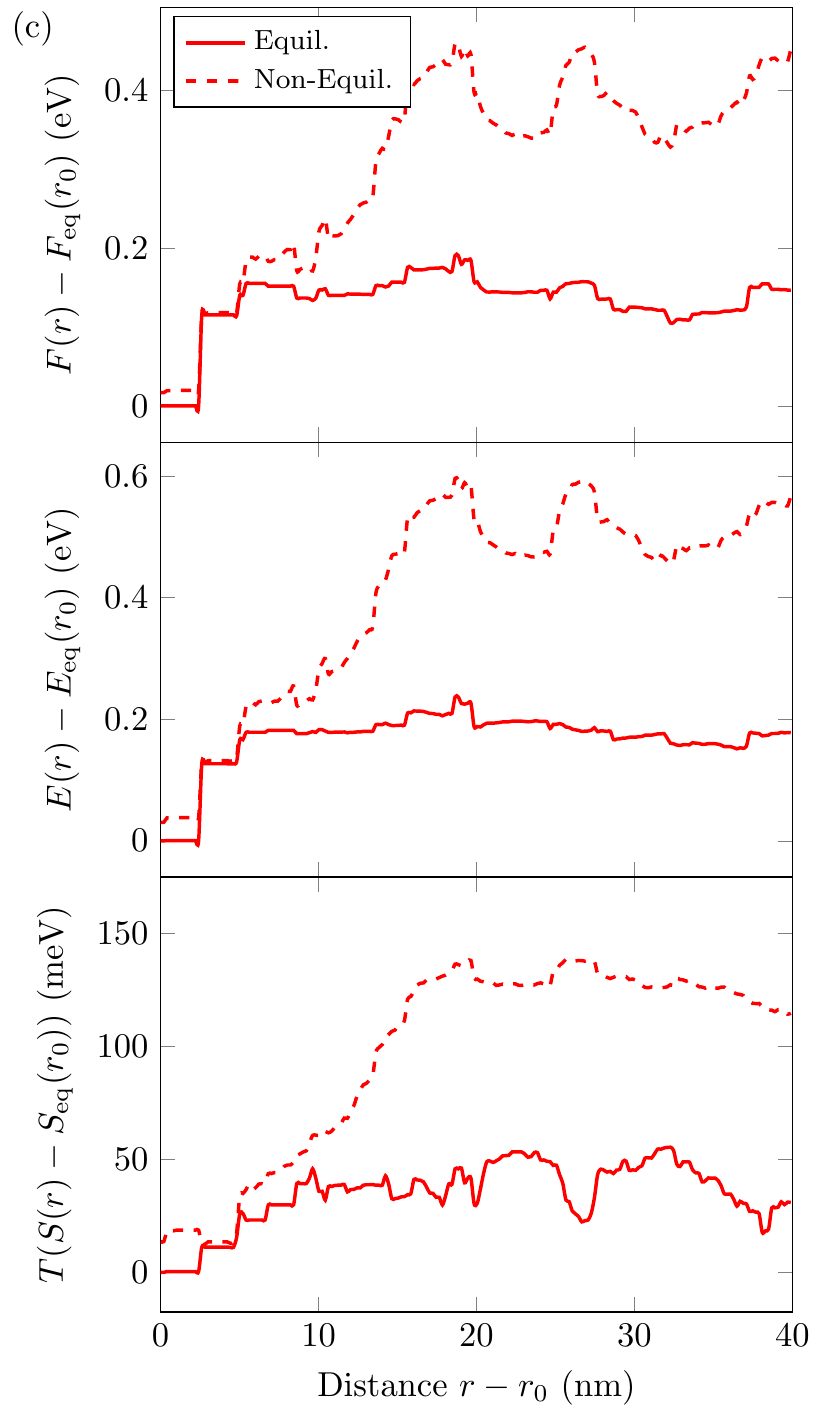}}
\end{minipage}
\caption{(a) Exemplary distribution of electron--hole pairs in the distance--energy phase space: donor excitons (green), CT states (red), and contact states (blue). (b) Time-dependence of the separation yield (Sep) and recombined states (GS) for the given network. (c) Distance dependence of the free energy $F(r)$, energy $E(r)$ and entropy contribution $TS(r)$. Solid lines represents the equilibrium values, dashed lines give the non-equilibrium results. All curves are normalized to the equilibrium value (subscript eq) at distance $r_0$.}
\label{fig:example}
\end{figure}

First, we present the capabilities of the developed method by calculating the distance dependence of the free energy ($F$) and its energy ($E$) and entropy contribution ($TS$) for a particular network of electron--hole pairs at zero field and $\sigma=\SI{50}{meV}$. Figure~\ref{fig:example}a shows the energy and the intrapair distance of each electron--hole pair (donor excitons, CT states, and contact states) in this particular network. CT and contact states both represent space-separated electron--hole pairs. The time evolution of separated and of recombined electron--hole pairs is visualized in Fig.~\ref{fig:example}b. This particular network shows a separation yield of \SI{83.3}{\percent}. The temporal evolution of the charge separation process and in particular "hot" and "cold" pathways were investigated in detail in Ref \citenum{jankovic2020energy}. Here, our focus is on the thermodynamic measures characterizing charge separation.

Figure~\ref{fig:example}c shows the equilibrium and non-equilibrium thermodynamic values as function of the intrapair distance $r$ for the given configuration. All curves are normalized to the equilibrium value at $r_0$, which is the smallest intrapair distance of all the CT states. The free energy $F$ and the energy $E$ are smallest at close intrapair distances. The one-dimensional model of the organic bilayer only shows very few CT states at short distance $r\leq\SI{5}{nm}$ which are all strongly bound, see Fig.~\ref{fig:example}a. The equilibrium energy increases by up to \SI{230}{meV} and remains roughly constant for distances beyond $r-r_0\geq\SI{5}{nm}$. The equilibrium free energy shows a similar trend, while being up to \SI{50}{meV} lower than the equilibrium energy. The difference is due to entropy contribution, which increases by \SI{50}{meV} within \SI{10}{nm} and remains roughly constant for larger distances. The non-equilibrium theory shows significant deviations from equilibrium for all thermodynamic measures. The non-equilibrium free energy increases up to \SI{0.45}{eV} at $r-r_0\approx\SI{20}{nm}$ and remains roughly constant for larger distances. The energy of occupied states even reaches values of $\SI{0.6}{eV}$. In addition, the non-equilibrium Gibbs entropy shows three times the value of the equilibrium entropy. Especially, a large slope in the entropy for intrapair distances below \SI{20}{nm} can be observed. Such an increase in entropy with distance is supporting charge separation and highly desirable for electron--hole pairs to overcome their mutual Coulomb attraction.

If we consider electron--hole pairs under equilibrium assumption, we immediately impose an instant thermalization of electron--hole pairs at each time and every position. This, however, is a strong assumption as highlighted by the non-equilibrium theory. Our model shows that the thermodynamic measures of electron--hole pairs are not well captured by the equilibrium assumption. The results suggest that the photogenerated electron--hole pairs efficiently propagate through CT states of higher energy and reach contact states before thermalization towards energetic tail states can occur. This explains the significantly higher value in Gibbs entropy, as more CT states are available at higher energies with respect to tail states, and consequently increases the chance for efficient charge separation. In realistic 3D systems, the entropy may reach even higher values and may dominate the energy contribution \cite{gregg2011entropy,ChemRev.110.6736,JPhysChemLett.4.1707}.


\subsection{Role of the Energetic Disorder} 

A critical connection between the energetic disorder $\sigma$ and free energy has been predicted based on the assumption of equilibrated CT states \cite{hood2016entropy}. We proceed our analysis and study the role of the energetic disorder $\sigma$ on the free energy $F(r)$ for $\sigma\in\{ \SI{50}{meV},\,\SI{100}{meV},\,\SI{150}{meV} \}$ at zero field. All remaining parameters are kept constant. For each $\sigma$, we take an ensemble average over 256 configurations to gain a realiable statistics. Note that we focus on the regime $r-r_0 \leq \SI{20}{nm}$, which represents the Bjerrum length of localized point charges for the chosen permittivity $\epsilon_r = 3.0$. For larger distances, the free energy remains roughly constant (see e.g. Fig.~\ref{fig:example}c).

\begin{figure}[t!]
    \centering\sbox{\measurebox}{%
  \begin{minipage}[b]{.52\textwidth}
  \subfloat{\label{fig:figA2}\includegraphics[width=\textwidth]{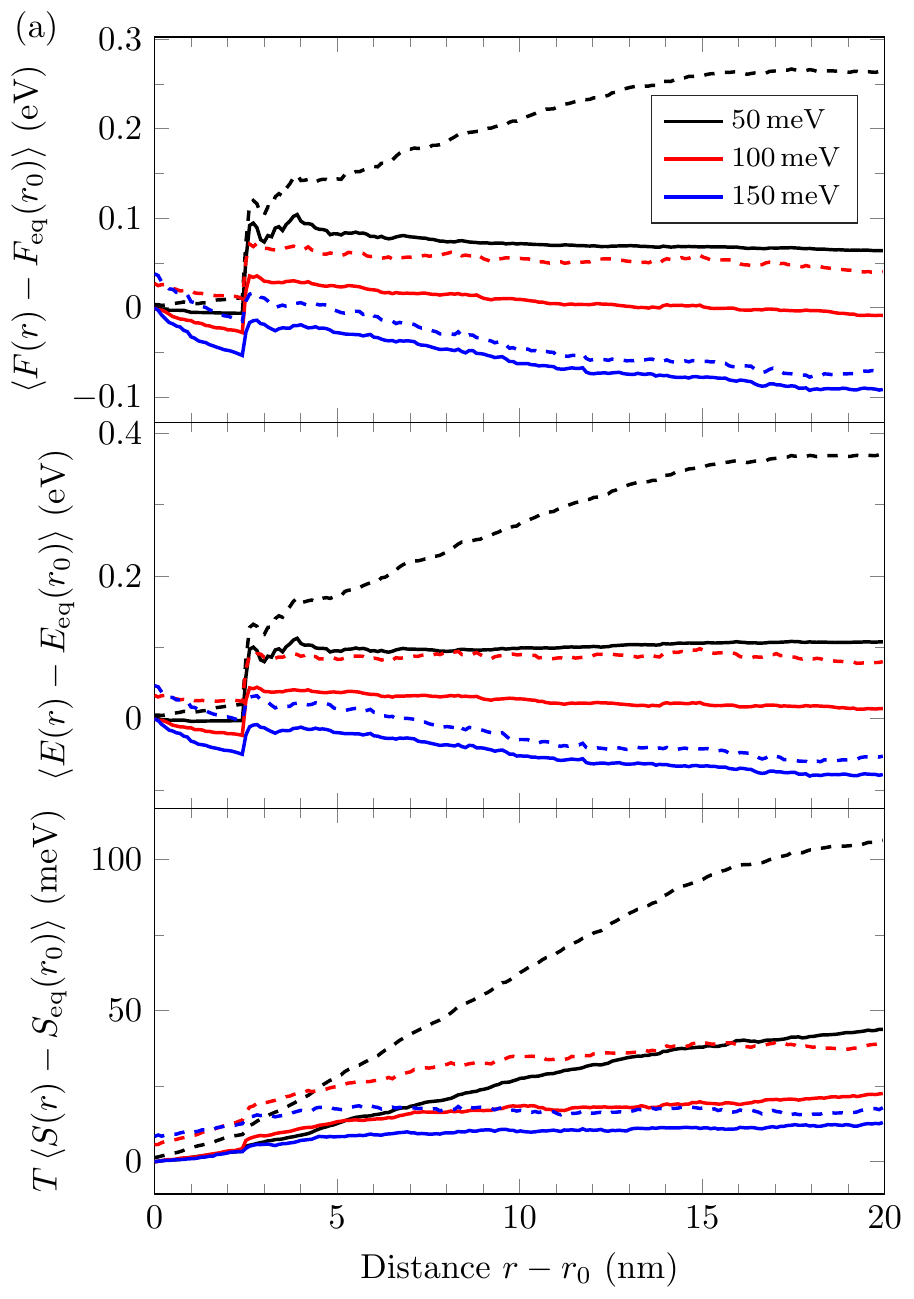}}
  \end{minipage}}
\usebox{\measurebox}
\begin{minipage}[b][\ht\measurebox][s]{.45\textwidth}
\centering
    \subfloat{\label{fig:figB2}\includegraphics[width=\textwidth]{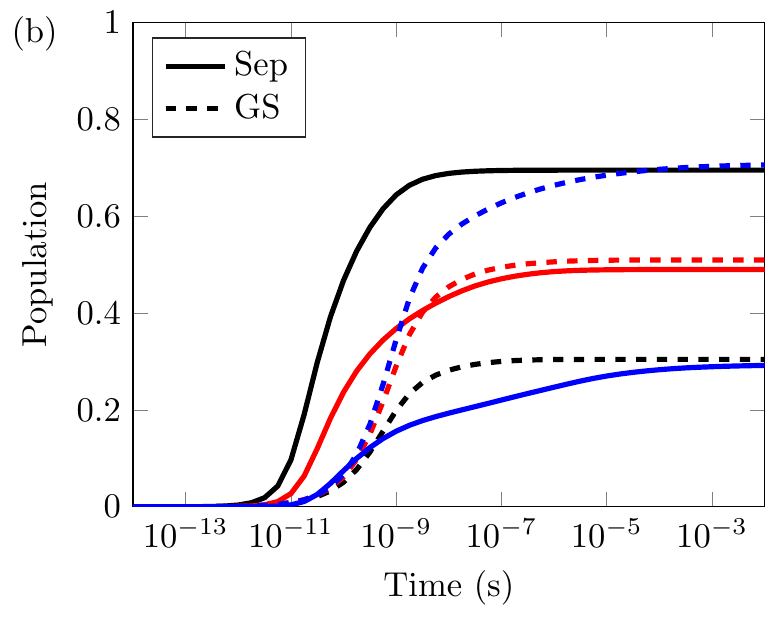}}
    \vfill
    \subfloat{\label{fig:figC2}\includegraphics[width=\textwidth]{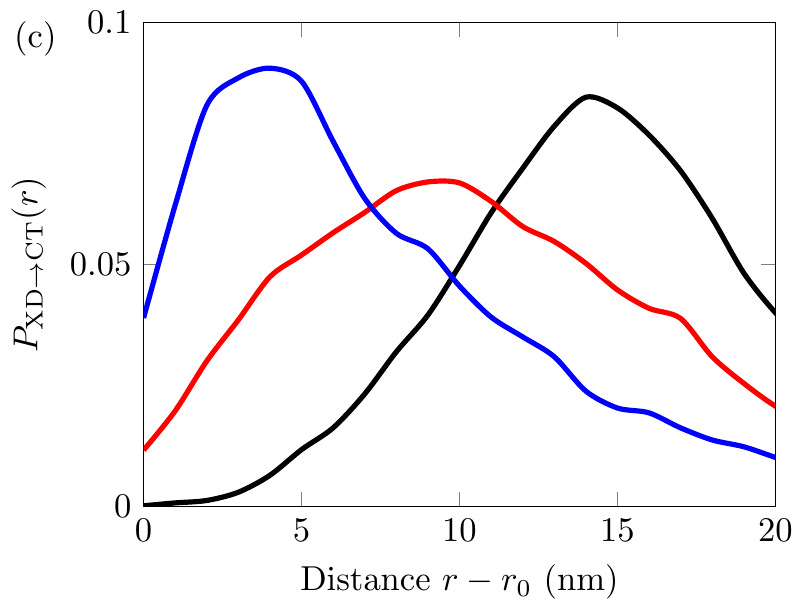}}
\end{minipage} 
\caption{(a) Distance dependence of the free energy $F(r)$, energy $E(r)$, and entropic contribution $TS(r)$ for different disorder $\sigma$: \SI{50}{meV} (black), \SI{100}{meV} (red), and \SI{150}{meV} (blue). $\left\langle\cdot\right\rangle$ labels the ensemble average over 256 configurations. $r_0$ gives the smallest distance of CT states in each configuration. Solid and dashed lines presents the equilibrium and non-equilibrium thermodynamic measures, respectively. All curves are normalized to the equilibrium value (subscript eq) at distance $r_0$. (b) Time-dependence of the separation yield (solid line) and the recombined states (dashed line). (c) Probability of donor exciton states (XD) to dissociate into CT states as function of the CT intrapair distances. All curves in (a) -- (c) show averages across 256 configurations.}
    \label{fig:td_disorder}
\end{figure}

Figure~\ref{fig:td_disorder}a shows the equilibrium and non-equilibrium free energy in dependence of the intrapair distance $r-r_0$. First, we analyze the results obtained under the assumption of equilibrated charge pair populations. The equilibrium free energy decreases strongly with increased $\sigma$. For $\sigma \geq \SI{100}{meV}$, $F$ even decreases for increased intrapair distance $r$. This observation matches previous results by Hood and Kassal \cite{hood2016entropy}. The decrease in free energy with $\sigma$ is mainly caused by a decrease in the average energy of the populated states during charge separation. The energy contribution shows the same distance-dependence as the free energy. The entropy contribution shows significant differences for different $\sigma$. At $\sigma=\SI{50}{meV}$, $TS$ increases monotonously with intrapair distance and reaches a value of \SI{45}{meV}. For larger $\sigma$, $TS$ is significantly reduced and reaches values of only \SI{25}{meV} (\SI{16}{meV}) at $\sigma = \SI{100}{meV}$ (\SI{150}{meV}). At high disorder, the entropy contribution remains roughly constant for $r - r_0 \geq \SI{5}{nm}$.
For low disorder of \SI{50}{meV}, significant differences between the equilibrium, $F_\mathrm{eq}$, and non-equilibrium free energy, $F_\mathrm{neq}$, are observable. $F_\mathrm{neq}$, computed from realistic populations of CT pairs, increases up to \SI{250}{meV} at $r-r_0 = \SI{20}{nm}$, while $F_\mathrm{eq}$ remains roughly constant for $r-r_0 \geq \SI{2.5}{nm}$. The entropy contribution shows a large increase up to \SI{108}{meV}, which explains the difference between the value in free energy and energy. For $\sigma \geq \SI{100}{meV}$, $E_\mathrm{neq}$ differs by less than \SI{50}{meV} from $F_\mathrm{neq}$ within the shown regime due to a low entropy contribution. At $\sigma=\SI{150}{meV}$, the entropy stays constant ($\approx \SI{20}{meV}$) within the full Bjerrum length. Entropy contributions which do not change with intrapair distance have been observed for 1D-systems without energetic disorder\cite{gregg2011entropy}, but also for 2D-systems with high energetic disorder\cite{hood2016entropy}.

Figure~\ref{fig:td_disorder}b visualizes the time dependence of the population of contact states, which is representing separated electron--hole pairs, and the amount of recombined states for different $\sigma$. With increased $\sigma$, the charge separation yield decreases from \SI{69.5}{\percent} ($\sigma =\SI{50}{meV}$) to \SI{49.0}{\percent} ($\sigma = \SI{100}{meV}$) and \SI{29.3}{\percent} ($\sigma = \SI{150}{meV}$). In addition, the time scale on which charge separation takes place increases by several orders of magnitude for larger $\sigma$. At high $\sigma$, electron--hole pairs thermalize within the DOS and populate the tail state. Equilibrated electron--hole pairs can only separate by propagation through low-energy states. However, the density of tail states does not increase significantly with intrapair distance. Large $\sigma$ helps to initially separate charge carriers in space, while low entropy slows down further charge transport tremendously. The lack of available states leads to high recombination losses and can promote non-geminate recombination -- which is neglected in this study -- as a further limitation of charge separation. 

To understand the origin of the efficient separation at low $\sigma$ despite the presence of the large free energy barrier, we analyze the dissociation of donor excitons (XD) into CT states by calculating the quantity
\begin{equation}
    P_{\mathrm{XD} \rightarrow \mathrm{CT}} (r) = \left(\sum_{x\in\mathrm{XD}, x'\in\mathrm{CT}} g_{x}w_{x'x}\right)^{-1} \sum_{x\in\mathrm{XD}, \left\lbrace x'\in\mathrm{CT} | r_{x'} \in\left[r - b/2, r + b/2\right)\right\rbrace} 
    g_{x} w_{x'x} 
\end{equation}
which gives the probability to dissociate a photoexcited donor state into a CT state with intrapair distance $r$, see Figure~\ref{fig:td_disorder}c. We see that with rising $\sigma$, the probability shifts to shorter distances, i.e. there is a higher chance to dissociate into a strongly-bound space-separate state with small intrapair distance. At $\sigma=\SI{50}{meV}$, donor excitons have a high probability to dissociate into CT states with large intrapair distance, which are comparable to "hot" CT states \cite{bassler2015hot,felekidis2020role}. Electron--hole pairs at distances larger than \SI{5}{nm} see a strong reduction in free energy barrier (see Figure~\ref{fig:td_disorder}a), while only strongly bound electron--hole pairs with $r\leq\SI{5}{nm}$ face a large barrier. Suppose the system starts in CT state with smallest distance $r_0$, i.e. in a strongly bound CT state. In that case, the thermodynamic measures of the electron--hole pairs during charge separation can be well described by the equilibrium populations, see Fig.~S2, Supporting Information. Thus, considering possible long-range dissociation of donor excitons into "hot" CT states is fundamental to accurately capture the underlying thermodynamics of charge separation in OSCs.

One of the main observations we can extract from Fig.~\ref{fig:td_disorder} is that the equilibrium theory is a good assumption for organic semiconductors with high energetic disorder. To achieve highly efficient OSCs, however, typically low $\sigma$ are required \cite{wu2020exceptionally, liu2020high}. High $\sigma$ hinder charge transport \cite{pasveer2005unified} and strongly reduce attainable open-circuit voltage \cite{blakesley2011relationship, kaiser2019kinetic}. For high $\sigma$, electron-hole pairs lose energy while moving towards the contact sites, matching previous reports \cite{melianas2015photo}. In contrast, the rather delocalized electron--hole pairs at $\sigma=\SI{50}{meV}$ do not thermalize and do not lose much less energy before being extracted at the contact sites. Thus, we emphasize that a non-equilibrium description is necessary to understand the role of the free energy for charge separation in efficient OSCs.

\subsection{Role of Charge Localization} 

\begin{figure}[ht!]
    \centering\sbox{\measurebox}{%
  \begin{minipage}[b]{.52\textwidth}
  \subfloat{\label{fig:figA3}\includegraphics[width=\textwidth]{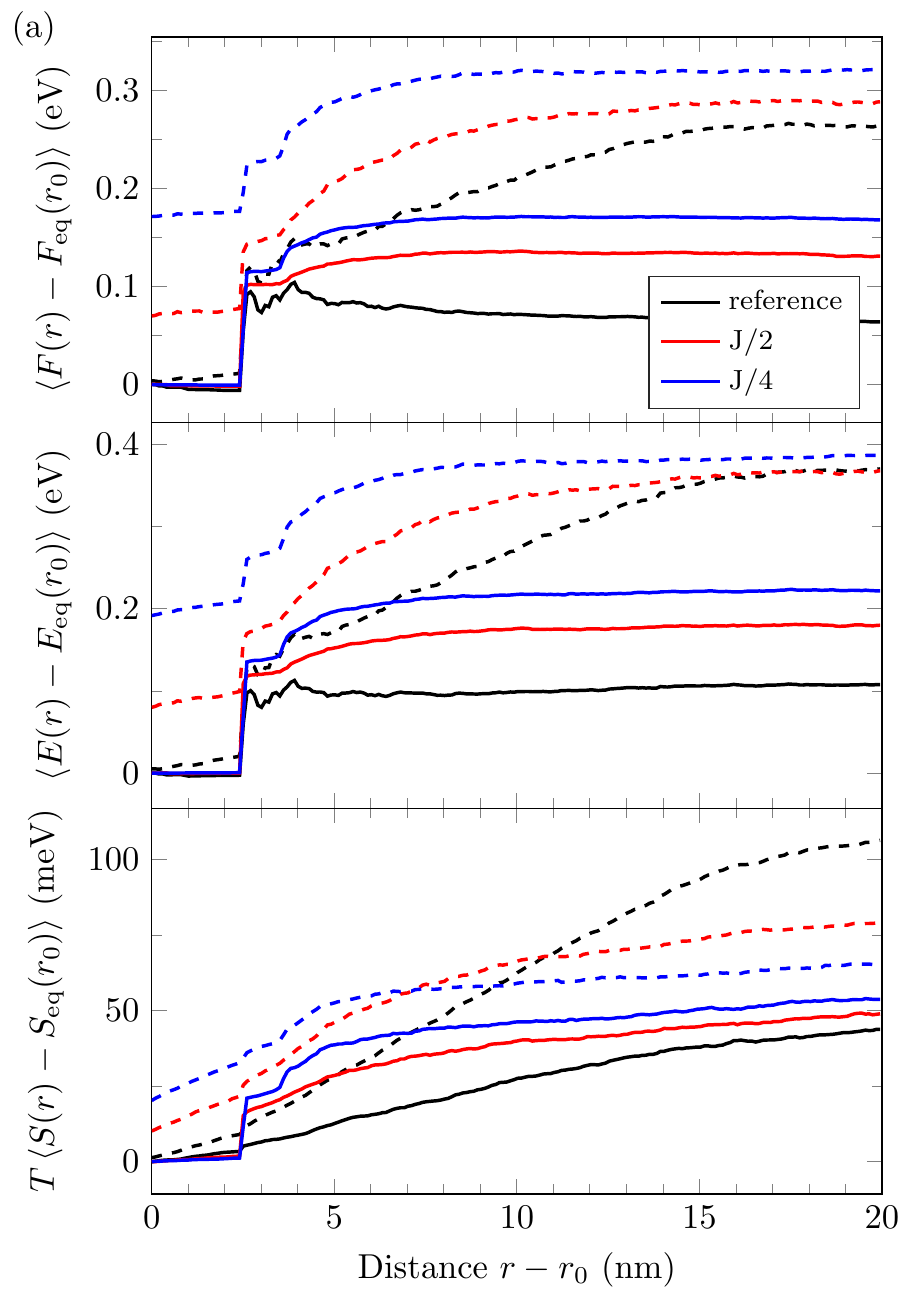}}
  \end{minipage}}
\usebox{\measurebox}
\begin{minipage}[b][\ht\measurebox][s]{.45\textwidth}
\centering
\subfloat{\label{fig:figB3}\includegraphics[width=\textwidth]{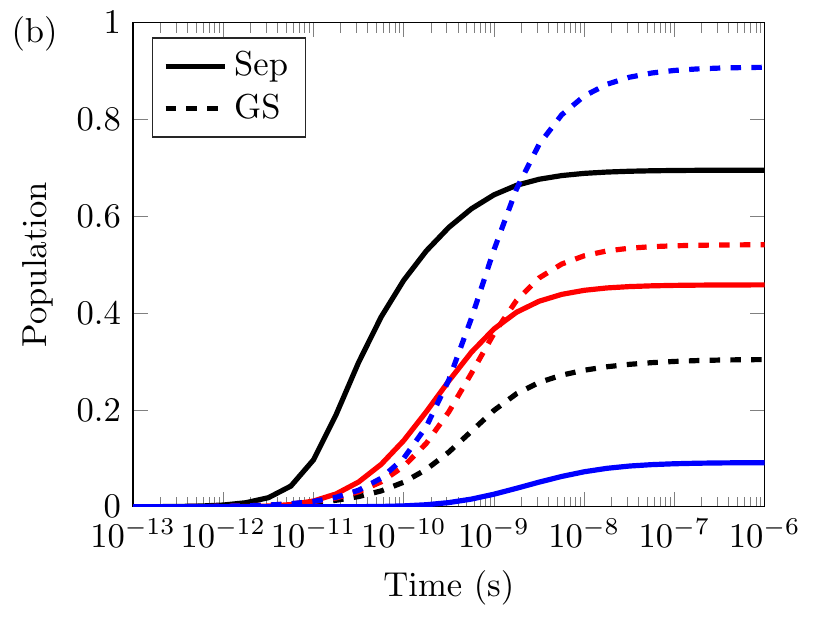}}
    \vfill
    \subfloat{\label{fig:figC3}\includegraphics[width=\textwidth]{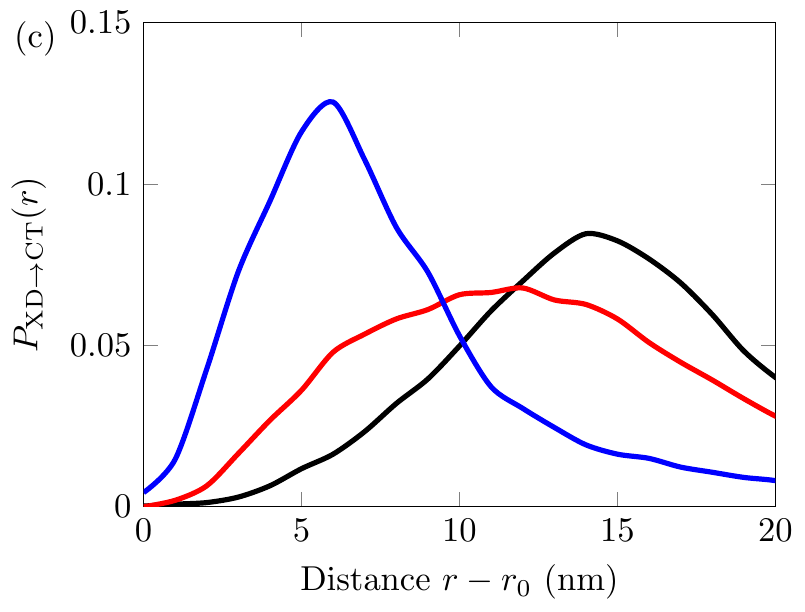}}
\end{minipage} 
\caption{(a) Distance dependence of the free energy $F(r)$, energy $E(r)$, and entropic contribution $TS(r)$ for different coupling integrals: reference (black), J/2 (red), and J/4 (blue). The average $\left\langle\cdot\right\rangle$ labels the ensemble average over 256 configurations. $r_0$ gives the smallest distance of CT states in each configuration. Solid line presents the equilibrium prediction, dashed lines give the non-equilibrium results. All curves are normalized to the equilibrium value (subscript eq) at distance $r_0$. (b) Time-dependence of the separation yield (solid line) and the recombined states (dashed line). (c) Probability of donor exciton states (XD) to dissociate into CT states as function of the CT intrapair distances. All curves in (a) -- (c) show averages across 256 configurations.}
    \label{fig:td_localization}
\end{figure}

Previous studies highlight the importance of delocalization in the separation of photogenerated electron--hole pairs in OSCs \cite{jankovic2018combination,kahle2018does,athanasopoulos2019binding, felekidis2020role}. To study the role of delocalization in the thermodynamics of the charge separation process, we increase the localization of CT states by scaling all transfer integrals $J_{A/D,0/1}^{c/v}$ (see Fig.~S1, Supporting Information) with a factor of $1/2$ and $1/4$. Figure~\ref{fig:td_localization}a shows the distance dependence of the equilibrium and non-equilibrium free energy and its energy and entropy contributions. According to both descriptions, the free energy increases with localization. Again, the energy follows the free energy closely. For $J/2$ and $J/4$, reduced separation yields of $45.9\%$ and $9.1\%$, respectively, are observed (see Fig.~\ref{fig:td_localization}b). This agrees with the increase in the free energy and energy barrier with larger localization. Interestingly, the non-equilibrium free energy deviates strongly from the equilibrium calculations for $J/2$ and $J/4$. At large distances, the non-equilibrium energies, $E_\mathrm{neq}$, for different localization approach the same value of $\approx\SI{380}{meV}$. In contrast, significant differences in $E_\mathrm{neq}$ are observed for short distances; strongly localized electron--hole pairs $J/4$ ($J/2$) occupy states with an energy of \SI{0.2}{eV} (\SI{0.1}{eV}) above equilibrium.  

The site occupation of electron--hole pairs at the interface determines the measured emission spectra by electro- and photoluminescence studies of donor-acceptor blends, which are frequently related to the open-circuit voltage in OSCs.\cite{vandewal2010relating, rau2007reciprocity} Our non-equilibrium results point out that, especially for strong localization, non-equilibrium site occupations of electron--hole pairs at the interface need to be considered. For large localization, the probability of donor excitons dissociating into a CT state with small intrapair distance is high (see Fig.~\ref{fig:td_localization}c). The state population at the interface, following exciton dissociation, is strongly impacted by the competition between thermalization dynamics and recombination. The significant deviation from equilibrium indicates that electron--hole pairs with small intrapair distances do not fully thermalize before recombination occurs. This is in line with recent electroluminescence\cite{melianas2019nonequilibrium} and photoluminescence studies\cite{brigeman2018nonthermal}. Strictly assuming equilibrated states may lead to a misinterpretation of experimental photo-/electroluminescence data.

With increased localization, Gibbs entropy contribution deviates at short intrapair distances by up to \SI{25}{meV} from equilibrium. The weak coupling integrals result in many CT states with small intrapair distances. In addition, excitons dissociate preferably into CT states of short distance (see Fig.~\ref{fig:td_localization}c). The values of the entropy differ from the equilibrium prediction as electron--hole pairs tend to recombine before thermalization occurs. Interestingly, the change in non-equilibrium entropy for $r-r_0\leq\SI{5}{nm}$ is roughly equal for the different cases (see Fig.~S3, Supporting Information). At larger distances, the increase in entropy is higher with less localized states. For $J/4$, $S_\mathrm{neq}$ remains nearly constant. Thus, charge separation gets suppressed with increased localization.  

In contrast to previous theoretical studies \cite{gluchowski2018increases}, our results emphasize that the free energy decreases with delocalization. In Ref.~\citenum{gluchowski2018increases}
the free energy was analyzed for electron-hole distances of less than \SI{4}{\nano\meter}. An increase in coupling reduces the amount of electron--hole pairs with small intrapair distance (cf. Ref.~\citenum{gluchowski2018increases}, Fig.~5), while only a few trap states in the tail of the DOS remain. This may explain the predicted increase in free energy. However, efficient OSCs are characterized by efficient long-range exciton separation, which leads to delocalized CT states of large initial mean distance \cite{kahle2018does, bakulin2012role, grancini2013hot, felekidis2020role} and consequently reduces the relevance of strongly-bound CT states. This comes closer to our model (see e.g. Fig.~\ref{fig:td_localization}c). 

\subsection{Transient Thermodynamic Free Energy}

\begin{figure}[t!]
    \centering
    \includegraphics[width=0.48\textwidth]{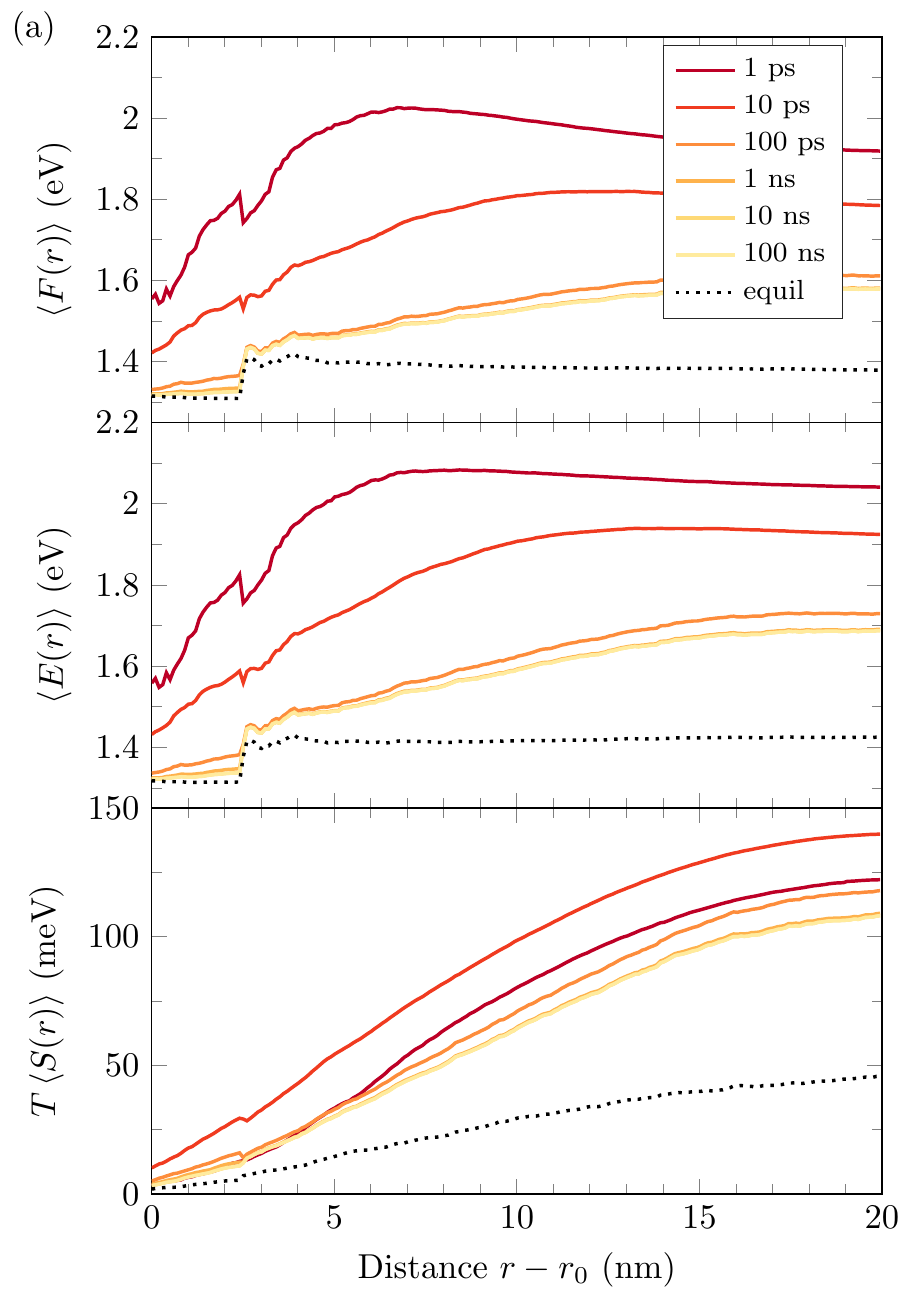}
    \includegraphics[width=0.48\textwidth]{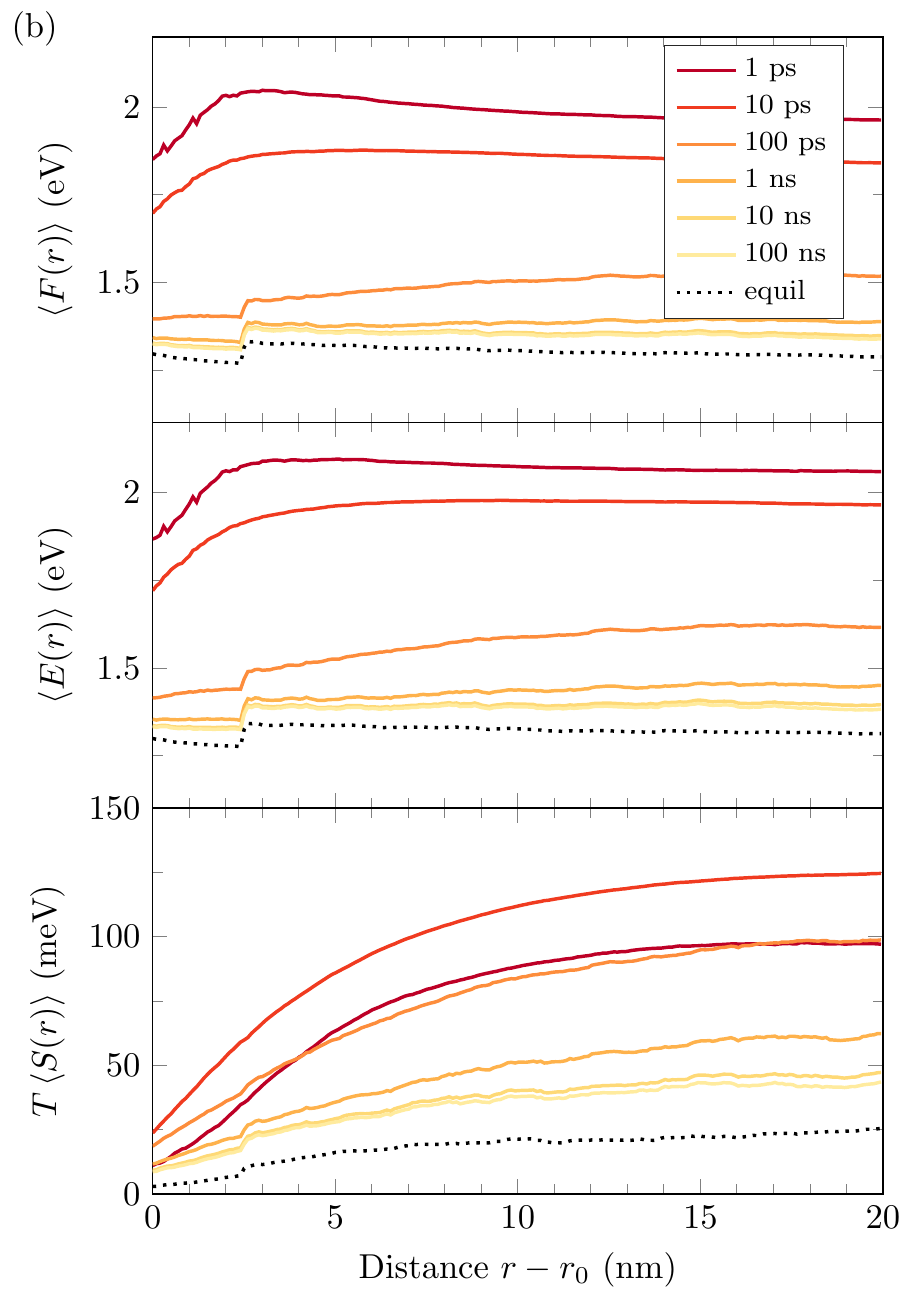}
\caption{Temporal evolution of the free energy $F(r)$, energy $E(r)$, and entropy contribution $TS(r)$ as function of the electron--hole intrapair distance for (a) $\sigma=\SI{50}{meV}$ and (b) $\sigma=\SI{100}{meV}$. The average $\left\langle\cdot\right\rangle$ labels the ensemble average over 256 configurations. $r_0$ gives the smallest distance of CT states in each configuration.}
    \label{fig:time_dep}
\end{figure}

So far, we only considered the free energy of electron--hole pairs during charge separation based on the steady-state populations. In contrast to previous equilibrium description, a salient feature of our methodology is that it allows us to study the time-dependence of the free energy and its energy and entropy contributions. Figure~\ref{fig:time_dep} shows the time evolution of the free energy for $\tau\in\left[\SI{1}{\pico\second},\dots,\SI{100}{\nano\second}\right]$ for a disorder of (a) $\sigma=\SI{50}{meV}$ and (b) $\sigma=\SI{100}{meV}$ (see Figs.~S4 and S5, Supporting Information, for other values of model parameters).

First, we analyze the temporal evolution of the free energy for $\sigma=\SI{50}{meV}$. The average energy $\left\langle E(r) \right\rangle$ decreases significantly with time, which can be interpreted as the relaxation of electron--hole pairs within the given density of CT states. The steady state is reached on a nanosecond time scale, which is representative of CT pair recombination and which represents the longest time scale in the problem. The free energy appears to follow the trend of the average energy. Interestingly, the entropy contribution $T\left\langle S(r)\right\rangle$ increases at early simulation times of up to \SI{10}{\pico\second}. On such short timescales, electron--hole pairs relax to the low-energy tail of the DOS, as was concluded in Ref.~\citenum{jankovic2020energy}, but their distribution is still far away from steady-state. This is so because our non-equilibrium thermodynamic measures reflect both the phonon-assisted state connectivity, as well as the finite lifetime of CT pairs, which jointly determine the charge separation routes that can actually be realized. For longer timescales, the maximum entropy contribution decreases from $\approx\SI{140}{meV}$ to $\approx\SI{110}{meV}$. 

For $\sigma=\SI{100}{meV}$, see Fig.~\ref{fig:time_dep}b, free energy and its energy contribution show the same behavior. Note that the timescale on which the steady state is established is somewhat longer, tens of nanoseconds, and steady-state values get closer to equilibrium. We have shown that the entropy contribution at large disorder strongly decreases. This can also be seen for ns-timescales and larger. However, at $\tau\leq\SI{1}{ns}$, high entropy terms of $\approx\SI{100}{meV}$ -- far beyond the maximum value of \SI{25}{meV} in equilibrium -- are observed. In addition, the timescales of entropy changes increase with $\sigma$ (see also Fig.~S5, Supporting Information) which can be related to a slow relaxation dynamics of electron--hole pairs. 

We can explain the evolution of the thermodynamic free energy by considering the relaxation dynamics of electron--hole pairs in a Gaussian DOS.\cite{baranovskii2014theoretical} The electron--hole pairs start far from equilibrium, while the average energy is high and only a few states are accessible, i.e. high energy and low entropy. During relaxation, the energy decreases while many more states become accessible near the maximum of the DOS. Once fully thermalized, the electron--hole pairs mainly occupy tail states of the DOS. Only a few states are left that can be accessed without an additional driving force while reaching states of higher energy requires an additional electric field. Relating these results to the work of Gregg \cite{gregg2011entropy}, we see that the number of accessible states changes with time by the relaxation of electron--hole pairs in the available DOS. This observation also allows us to specify a time scale for the most efficient separation of "hot" electron--hole pairs within $\leq \SI{1}{\nano\second}$. Our observations are consistent with the results of Ref.~\citenum{PhysRevLett.114.247003}, according to which the entropic gain enables non-equilibrated CT pairs to overcome the Coulomb barrier on a very short time scale, in contrast to equilibrated CT pairs that are bound to recombine due to the lack of accessible states.

While most existing studies rely on the characterization of steady-state electro- and photoluminescence \cite{brigeman2018nonthermal,melianas2019nonequilibrium,vandewal2010relating}, experimental techniques such as ultrafast spectroscopy \cite{moore2020ultrafast,tsokkou2020excited} or time-resolved X-ray photoemission spectroscopy\cite{roth2021direct} can provide insight into the dynamics of excited states on sub-picosecond timescales. Recent studies pointed out the relevance of ultrafast charge separation in the picosecond regime in non-fullerene acceptor-based OSCs.\cite{tamai2017ultrafast,wang2020charge} We strongly believe that our presented study can be of high relevance as a complementary tool to transient spectroscopy techniques to understand the role of the thermodynamic free energy on the charge separation as it is providing hitherto inaccessible quantification on the temporal evolution of the individual components of the thermodynamic free energy.

\section{Conclusion}
\label{sec:conclusion}

In conclusion, we have presented a theoretical description of the non-equilibrium free energy barrier for charge separation in organic solar cells based on a 1D-model Hamiltonian. Our methodology combines the phonon-assisted kinetics of electron--hole pairs with the thermodynamic free energy based on the concept of stochastic thermodynamics. The presented method allows calculating the free energy and Gibbs entropy for any system of Markovian dynamics. The role of the energetic disorder and delocalization has been analyzed based on the steady-state populations of space-separated states. Our analysis reveals significant deviations from equilibrium in the free energy and the entropy for delocalized electron--hole pairs at small energetic disorder, representing efficient OSCs. In systems of large energetic disorder, the steady-state site occupation of electron--hole pairs can be well described as equilibrated. Localized electron--hole pairs with short intrapair distances show significant non-equilibrium site occupation, explaining previous observations from photoluminescence measurements. Our results emphasize that both a large Gibbs entropy and large initial separation are required to achieve an efficient separation, while a decrease with distance in free energy values does not necessarily correlate with an enhanced separation. We expect the entropy contribution to be of even higher relevance within realistic 3D systems. One of the most relevant novelties of the presented framework is the ability to connect the thermodynamic free energy with the electron--hole pair dynamics in terms of the temporal evolution of the free energy. We show that the entropy significantly changes within the relevant timescales of charge separation. Sub-ns timescales provide largest Gibbs entropy, which may explain the efficient separation of "hot" CT states on such short timescales, supporting recent studies on efficient non-fullerene solar cells. Thermalization leads to a substantial reduction in entropy and reduces the effective DOS accessible for electron--hole pairs during charge separation. Our non-equilibrium thermodynamic description opens hitherto inaccessible insight into physical processes and may provide a framework in combination with time-dependent experiments for future studies.

\section*{Acknowledgement}
W.K. and A.G. acknowledge the TUM International Graduate School of Science and Engineering (IGSSE) and Germany's excellence cluster \textit{e}-conversion by the German Research Foundation (Deutsche Forschungsgemeinschaft, DFG) for funding. N.V. and V.J. acknowledge funding by the Institute of Physics Belgrade, through the grant by the Ministry of Education, Science, and Technological Development of the Republic of Serbia. Numerical computations were partially performed on the PARADOX-IV supercomputing facility at the Scientific Computing Laboratory, National Center of Excellence for the Study of Complex Systems, Institute of Physics Belgrade.

\section*{Supporting Information}
Description of the model Hamiltonian and of the transition rate derivation of the organic bilayer; model parameters used in computations; numerical solution procedure of the Master equation and transient thermodynamic measures; transient thermodynamic measures for different delocalization and disorder.

\bibliography{references}




\end{document}


\maketitle
\pagebreak

\section*{Model Hamiltonian}

The organic bilayer is modeled using a 1-dimensional model Hamiltonian which has been presented in previous publications \cite{jankovic2018combination}. Here, we briefly outline the scheme of the model and the used parameter set. 

\begin{figure}
    \centering
    \includegraphics[width=12cm]{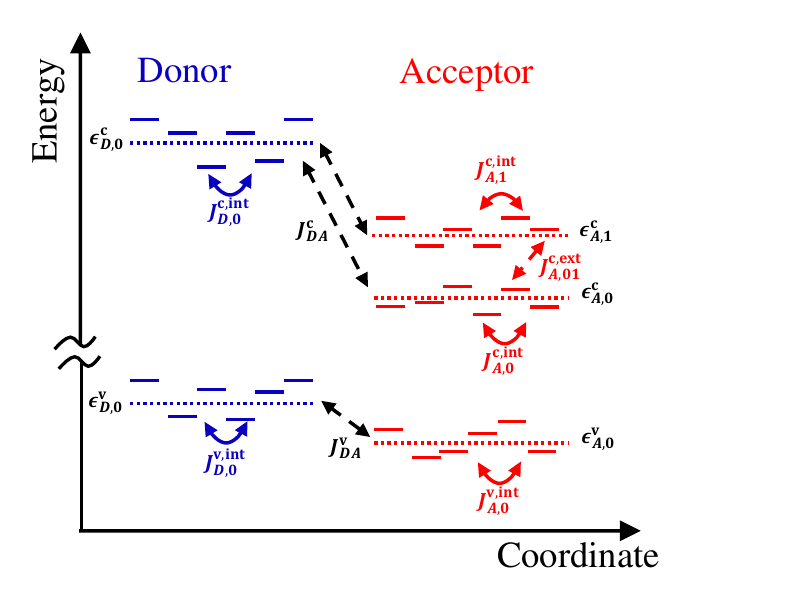}
    \caption{Schematic of the model Hamiltonian of the organic donor:acceptor bilayer. The dotted lines highlight the different average on-site energies. Solid lines represent the actual on-site energies which vary from the average on-site energies due to the static energetic disorder. The different transfer integrals within the donor (D) and acceptor (A) as well as the coupling between the donor and acceptor (DA) are highlighted.}
    \label{fig:model_scheme}
\end{figure}

Figure~\ref{fig:model_scheme} visualizes the system setup of the 1D-bilayer. The one-dimensional system consists of $2N$ sites located on a lattice with spacing $a$. Sites $0,\dots,N-1$ represent the donor, sites $N,\dots,2N -1$ represent the acceptor of the bilayer. We account for energetically disordered HOMO and LUMO levels of the donor and acceptor with disorder $\sigma$. For the acceptor, we also include the LUMO+1 level. Transfer integrals are accounted for between all neighbor sites with different magnitude in the HOMO / LUMO levels of the acceptor and donor materials itself, as well as between the donor and the acceptor. The contact region, in which electrons and holes are mainly localized when they are fully separated, consists of sites $0,\dots,l_c-1$ in the donor part (holes) and sites $2N-l_c,\dots,2N-1$ in the acceptor part (electrons) of the bilayer.

The model Hamiltonian is accounts for charge interaction, $\hat{H}_c$, the phonon bath, $\hat{H}_p$, and the interaction of charge carriers with the phonons, $\hat{H}_{c-p}$:

\begin{equation}
    \hat{H} = \hat{H}_c + \hat{H}_p + \hat{H}_{c-p}
\end{equation}
The charge contributions are modeled by
\begin{align}
    \hat{H}_c &= \sum_{i\beta_i, j\beta_j} \epsilon_{(i\beta_i),(j,\beta_j)}^{c}c_{i\beta_i}^\dagger c_{j\beta_j} - \sum_{i\alpha_i, j\alpha_j} \epsilon_{(i\alpha_i),(j,\alpha_j)}^{v}d_{i\alpha_i}^\dagger d_{j\alpha_j} \\
    &+ \frac{1}{2} \sum_{i\beta_i, j\beta_j} V_{ij} c_{i\beta_i}^\dagger c_{j\beta_j}^\dagger c_{j\beta_j} c_{i\beta_i}
    + \frac{1}{2} \sum_{i\alpha_i, j\alpha_j} V_{ij} d_{i\alpha_i}^\dagger d_{j\alpha_j}^\dagger d_{j\alpha_j} d_{i\alpha_i} \nonumber\\
    &- \sum_{i\beta_i, j\alpha_j} V_{ij} c_{i\beta_i}^\dagger d_{j\alpha_j}^\dagger d_{j\alpha_j} c_{i\beta_i}\nonumber
\end{align}
The Fermi operators $c_{i\beta_i}^\dagger$ ($c_{i\beta_i}$) create (annihilate) an electron on site $i$ in single electron state $\beta_i$, $d_{i\alpha_i}^\dagger$ ($d_{i\alpha_i}$) create (annihilate) a hole on site $i$ in single hole state $\alpha_i$. The phonon bath is modeled by
\begin{equation}
    \hat{H}_p = \sum_{i\lambda}\hbar\omega_\lambda b_{i\lambda}^\dagger b_{i\lambda}
\end{equation}
with the Bose operators $b_{i\lambda}^\dagger$ ($b_{i\lambda}$), site $i$ and phonon mode $\lambda$. The interaction between the charges and the phonon bath is modeled by
\begin{equation}
    \hat{H}_{c-p} = \sum_{i\beta_i}\sum_{\lambda}g_{i\beta_i\lambda}^{c} c_{i\beta_i}^\dagger c_{i\beta_i}\left( b_{i\lambda}^\dagger + b_{i\lambda} \right) - \sum_{i\alpha_i}\sum_{\lambda}g_{i\alpha_i\lambda}^{v} d_{i\alpha_i}^\dagger d_{i\alpha_i}\left( b_{i\lambda}^\dagger + b_{i\lambda} \right)
\end{equation}
where for the coupling constants $g_{i\beta_i\lambda}^{c}$ and $g_{i\alpha_i\lambda}^{v}$ we assume $g_{i\beta_i\lambda}^{c}=g_{i\alpha_i\lambda}^{v}=g_\lambda$.

The Coulomb interaction is approximated by the Ohno potential
\begin{equation}
    V_{ij} = \frac{U}{\sqrt{1 + \left(\frac{r_{ij}}{r_0}\right)^2}}
\end{equation}
with $r_0=q^2/(4\pi\epsilon_0\epsilon_r U)$, the on-site Coulomb potential $U$, $r_{ij}$ the distance between sites $i$ and $j$, the permittivity $\epsilon_r$, and the elementary charge $q$.

The coupling of electronic excitations to the phonon bath is assumed to be characterized (at every site) by the Ohmic spectral density with the exponential cutoff
\begin{equation}
    \mathcal{J}(E)=\sum_\lambda g_\lambda^2\:\delta(E-\hbar\omega_\lambda)=\eta E\:e^{-E/E_c}
\end{equation}
where $E_c$ denotes the cutoff energy, while $\eta$ measures the overall strength of the electron--phonon interaction. The polaron binding energy (in the single-site limit) is then $E_\mathrm{pol}=\eta E_c$. We compute the transition rate $w_{x'x}$ from excitonic state $|x\rangle$ to excitonic state $|x'\rangle$ by
\begin{equation}
\label{Eq:supp-tr-rates}
    w_{x'x}=\frac{2\pi}{\hbar}P_{x'x}\:\mathcal{J}\left(\left|E_{x'}-E_x\right|\right)n(E_{x'}-E_x)
\end{equation}
In Eq.~\ref{Eq:supp-tr-rates}, $P_{x'x}$ is the so-called spatial proximity factor, which summarizes the dependence of $w_{x'x}$ on spatial properties (e.g., wave function overlap, degree of delocalization) of states $|x\rangle$ and $|x'\rangle$,~\cite{jankovic2018combination} while $n(E)$ discriminates between energetically uphill and downhill transitions according to
\begin{equation}
    n(E)=\begin{cases}
    n_\mathrm{BE}(E),&E>0\\
    1+n_\mathrm{BE}(E),&E<0
    \end{cases}
\end{equation}
where $n_\mathrm{BE}(E)=\left[\exp(E/(k_BT))-1\right]^{-1}$ is the Bose--Einstein factor at phonon-bath temperature $T$.

\pagebreak
The values of the model parameters used in our calculations are summarized in Table~\ref{tab:parameterlist}. These are the default values assumed in the calculations, while some of these parameters were further varied to investigate the role of disorder and delocalization. The labels of different parameter sets used for such investigations are given in Table \ref{tab:table_sepyields}.

\begin{table}[]
    \caption{Values of model parameters used in computations. The parameters $\tau_0$ and $A_\text{A/D}$ determine the recombination lifetime of the states. The definition of these parameters is given in Ref. \citenum{jankovic2018combination}.}
        \begin{tabular}{|cc|}
        \hline 
        Parameter & Value \\ 
        \hline 
        $N$ & 30 \\ 
        $l_c$ & 11 \\ 
        $a$ (nm) & 1.0 \\ 
        $U$ (eV) & 0.65 \\ 
        $\epsilon_r$ & 3.0 \\ 
        $\epsilon_{D,0}^\text{c}$ (eV) & 2.63 \\ 
        $J_{D,0}^{\text{c,int}}$ (eV) & 0.1 \\
        $\epsilon_{D,0}^\text{v}$ (eV) & -0.3 \\ 
        $J_{D,0}^{\text{v,int}}$ (eV) & -0.15 \\
        $\epsilon_{A,0}^\text{c}$ (eV) & 1.565 \\ 
        $\epsilon_{A,1}^\text{c}$ (eV) & 1.865 \\ 
        $J_{A,0}^{\text{c,int}}$ (eV) & 0.05 \\
        $J_{A,1}^{\text{c,int}}$ (eV) & 0.025 \\
        $J_{A,01}^{\text{c,ext}}$ (eV) & 0.02 \\ 
        $\epsilon_{A,0}^\text{v}$ (eV) & -1.03 \\ 
        $J_{A,0}^{\text{v,int}}$ (eV) & -0.15 \\
        $J_{DA}^\text{c}$ (eV) & 0.1 \\
        $J_{DA}^\text{v}$ (eV) & -0.1 \\
        $\sigma$ (meV) & 50 \\ 
        $\eta$ & 1.5 \\ 
        $E_c$ (meV) & 10 \\ 
        $\tau_0$ (ps) & 400 \\ 
        $A_\text{A/D}$ & 0.5 \\ 
        $T$ (K) & 300 \\ 
        \hline 
        \end{tabular} 
    \label{tab:parameterlist}
\end{table}

\pagebreak
\section*{Numerical Solution of the Master Equation}

Here, we provide more details regarding the procedure to compute transient thermodynamic measures. The population $f_m(t)$ of the state $\ket{m}$ evolves according to the Master equation:
\begin{equation}
    \partial_t f_m(t) = \sum_{n \neq m} w_{mn} f_n(t) - \sum_{n \neq m} w_{nm} f_m(t) - \tau_m^{-1}f_m(t)\,,
\end{equation}
where $\tau_m$ is the recombination lifetime of state $\ket{m}$.
The population of the ground state (GS) satisfies:
\begin{equation}
    \partial_t f_{\mathrm{GS}}(t) = \sum_{m}\tau_{m}^{-1}f_m(t)\,.
\end{equation}
Contact states $c$ are assumed to be ideal. If a CT pair reaches a contact state, the charges are immediately collected into the external circuit. Thus, the population of contact states evolves as follows:
\begin{equation}
    \partial_t f_c(t) = \sum_{m} w_{cm} f_m(t)\,, 
\end{equation}
where $m$ labels all CT states. The system of rate equations can be expressed by a time-independent matrix $\widehat{W}$ in matrix form:
\begin{equation}
    \partial_t f_m(t) = \sum_{n} W_{mn}f_n(t)\,.
\end{equation}
If the initial condition $f_m(0)$ is known, we can solve the time evolution as
\begin{equation}
    f_m(t) = \left\langle m | \boldsymbol{f}(t)\right\rangle = \braket{m|\exp(\widehat{W}t)|\boldsymbol{f}(0)}\,.
\end{equation}
In our computations, $\boldsymbol{f}(0)=\boldsymbol{g}$, i.e., the initial condition is set by the generation rates $\boldsymbol{g}$ into donor exciton states, see Eqs.~2,~9~and~10 of the main text. As described in Ref. \citenum{jankovic2020energy}, we can express the solution of $f_m(t)$ in terms of the eigenvalues $\lambda_k$ and eigenvectors $c_k$ of the matrix $\widehat{W}$:
\begin{equation}
    f_m(t) = \sum_k \alpha_k \exp{\left(\lambda_k t\right)}c_{mk}\,,
\end{equation}
where the coefficients $\alpha_k$ can be found by solving the linear algebraic equations 
\begin{equation}
    f_m(0) = \sum_k \alpha_k c_{mk}\,.
\end{equation}
To obtain the non-equilibrium probability of being in state $m$ in the time interval $[0,\tau]$ during the separation of charge carriers, we take the time-average of probability $p_m(t)$, see Eq.~10 of the main text:
\begin{equation}
\label{Eq:f-m-neq-tau}
    f_{m}^\mathrm{neq}(\tau) = \tau^{-1} \int_{0}^{\tau} \mathrm{d}t f_m(t)\,.
\end{equation}
We consider the values of $\tau$ up to those when all CT states reached the contact states or recombined to the ground state. For a given $r$, the time-dependent non-equilibrium distribution $p_m^\mathrm{neq}(\tau)$ of CT states whose intrapair separation $r_m\in[r-b/2,r+b/2)$ is obtained from time-dependent populations $f_m^\mathrm{neq}(\tau)$ entering Eq.~\ref{Eq:f-m-neq-tau} as follows (cf. Eq.~6 of the main text)
\begin{equation}
    p_m^\mathrm{neq}(\tau)=f_m^\mathrm{neq}(\tau)\left(\sum_{ \left\lbrace m | r_m \in [r - b/2, r + b/2) \right\rbrace}f_m^\mathrm{neq}(\tau)\right)^{-1} \,.
\end{equation}

\section*{Additional Data}
\begin{table}[]
    \centering
    \begin{tabular}{lcc}\hline
        Parameter set & Properties & Separation Yield (\si{\percent}) \\\hline
        \textbf{default\_parameters} & $\sigma=\SI{50}{meV}$, default $J$, $F=0$ & 69.5 \\
        \textbf{sigma\_100meV} & $\sigma=\SI{100}{meV}$, default $J$, $F=0$ & 49.0 \\
        \textbf{sigma\_150meV} & $\sigma=\SI{150}{meV}$, default $J$, $F=0$ & 29.3 \\
        \textbf{all\_transfer\_less\_2} & $\sigma=\SI{50}{meV}$, $J/2$, $F=0$ & 45.9 \\
        \textbf{all\_transfer\_less\_4} & $\sigma=\SI{50}{meV}$, $J/4$, $F=0$ & 9.1 \\
        \hline
    \end{tabular}
    \caption{Separation yield averaged across 256 configurations for each parameter set.}
    \label{tab:table_sepyields}
\end{table}
\begin{figure*}[h!]
    \centering
    \includegraphics[width=8cm]{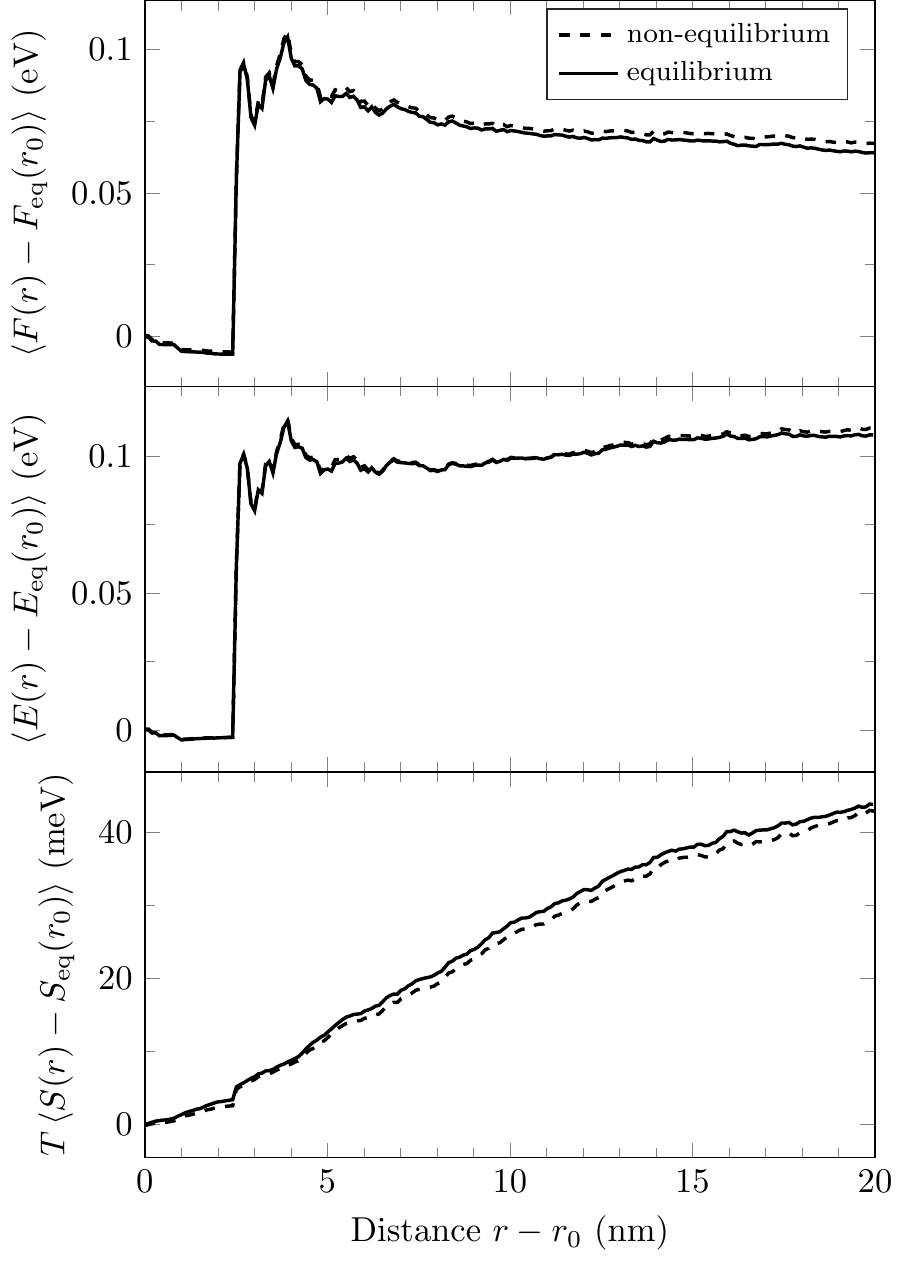}
    \caption{Distance dependence of the free energy $F(r)$, energy $E(r)$, and entropy contribution $TS(r)$ for parameter set \textbf{default\_parameters}. As initial state, we assume that only the space-separated state with smallest distance $r_0$ is populated. Solid (dashed) line presents predictions by equilibrium (non-equilibrium) method. All curves are normalized to the equilibrium value (subscript eq) at distance $r_0$.}
    \label{fig:sigma50_r0}
\end{figure*}
\begin{figure*}[h!]
    \centering
    \includegraphics[width=8cm]{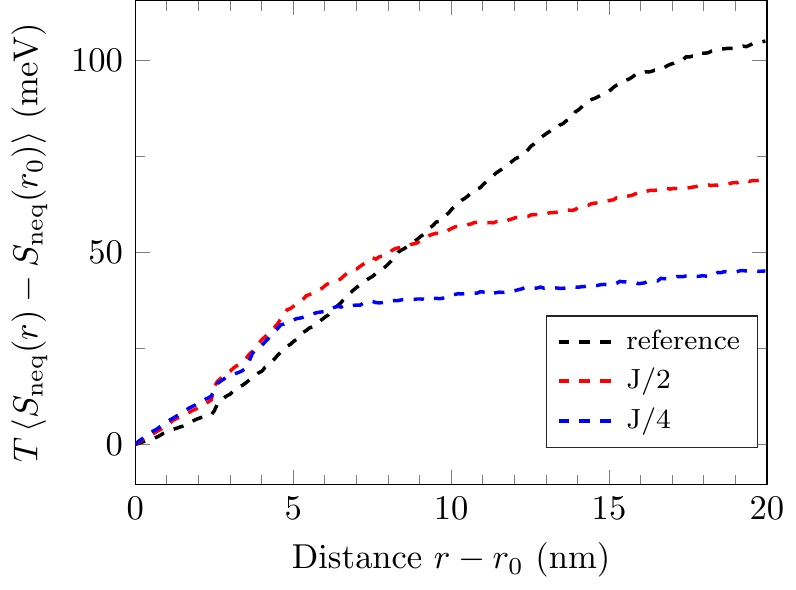}
    \caption{Non-equilibrium entropy for different localization - the parameter sets \textbf{default\_parameters}, \textbf{all\_transfer\_less\_2} and \textbf{all\_transfer\_less\_4}.}
    \label{fig:coupl_neq_entropy}
\end{figure*}
%
%
\begin{figure}
    \centering
    \includegraphics[width=8cm]{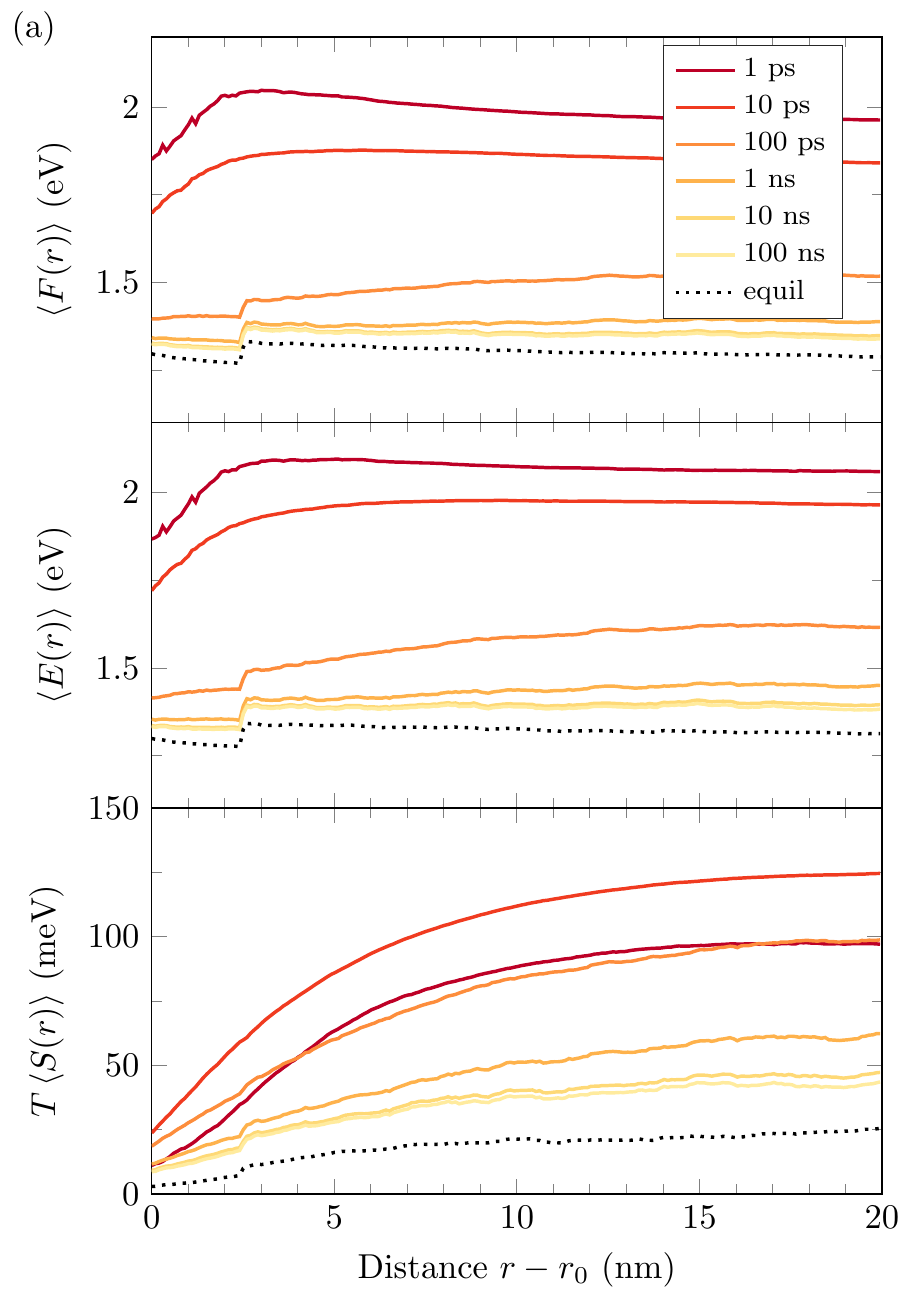}
    \includegraphics[width=8cm]{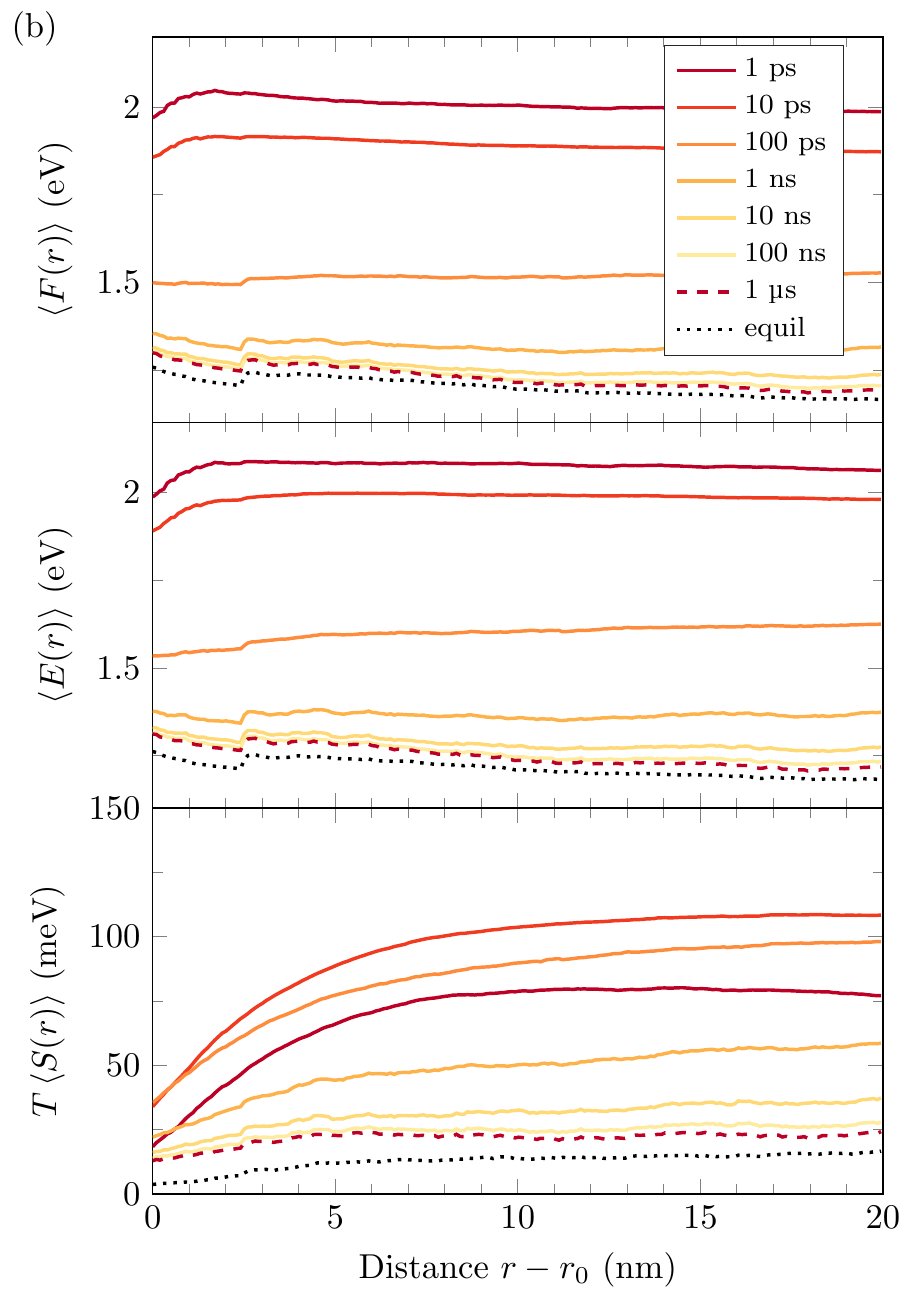}
    \caption{Time dependence of the free energy and its energy and entropy contribution for the parameter set (a) \textbf{sigma\_100meV} and (b) \textbf{sigma\_150meV}.}
    \label{fig:time_dep_disorder}
\end{figure}
\begin{figure}
    \centering
    \includegraphics[width=8cm]{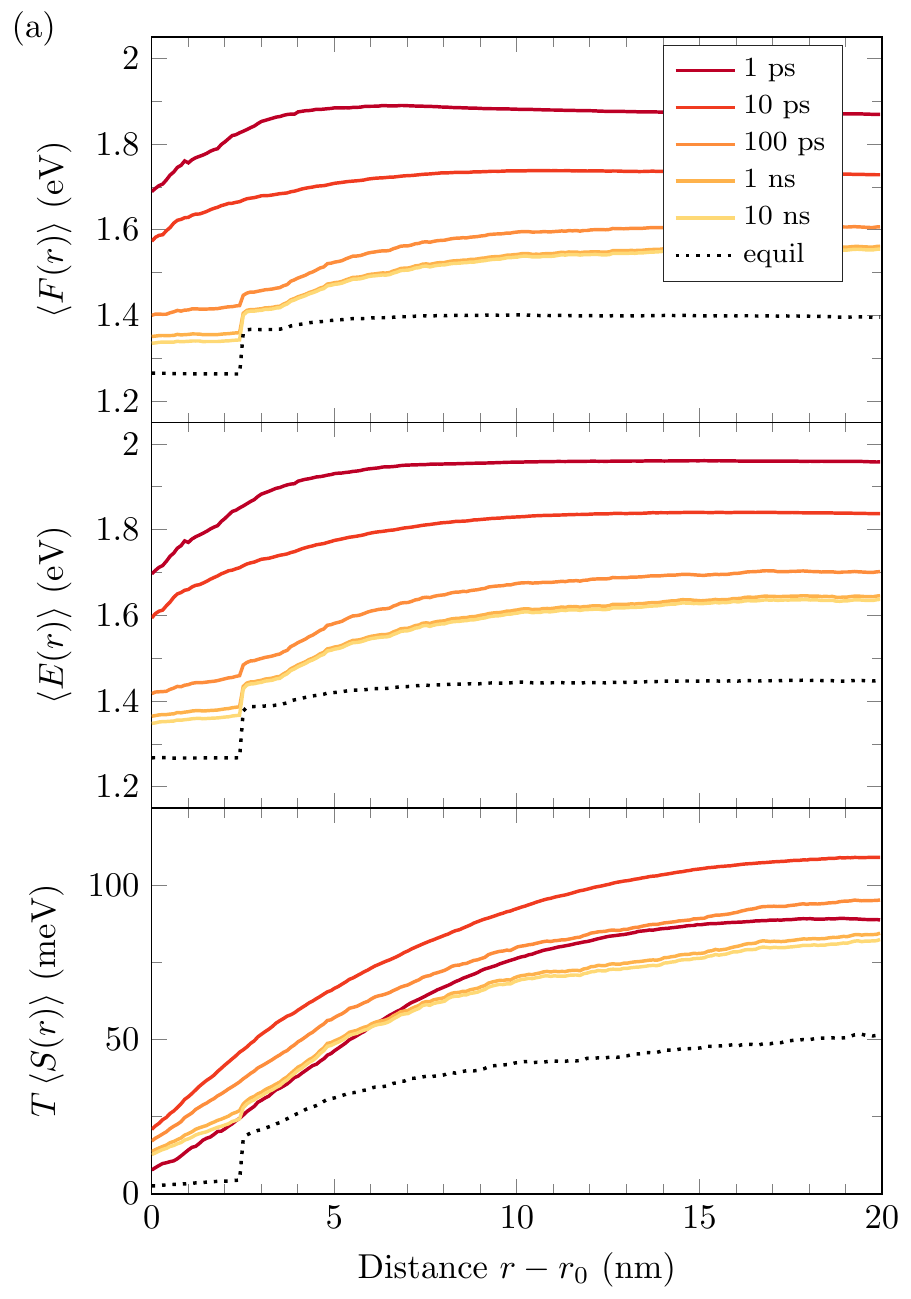}
    \includegraphics[width=8cm]{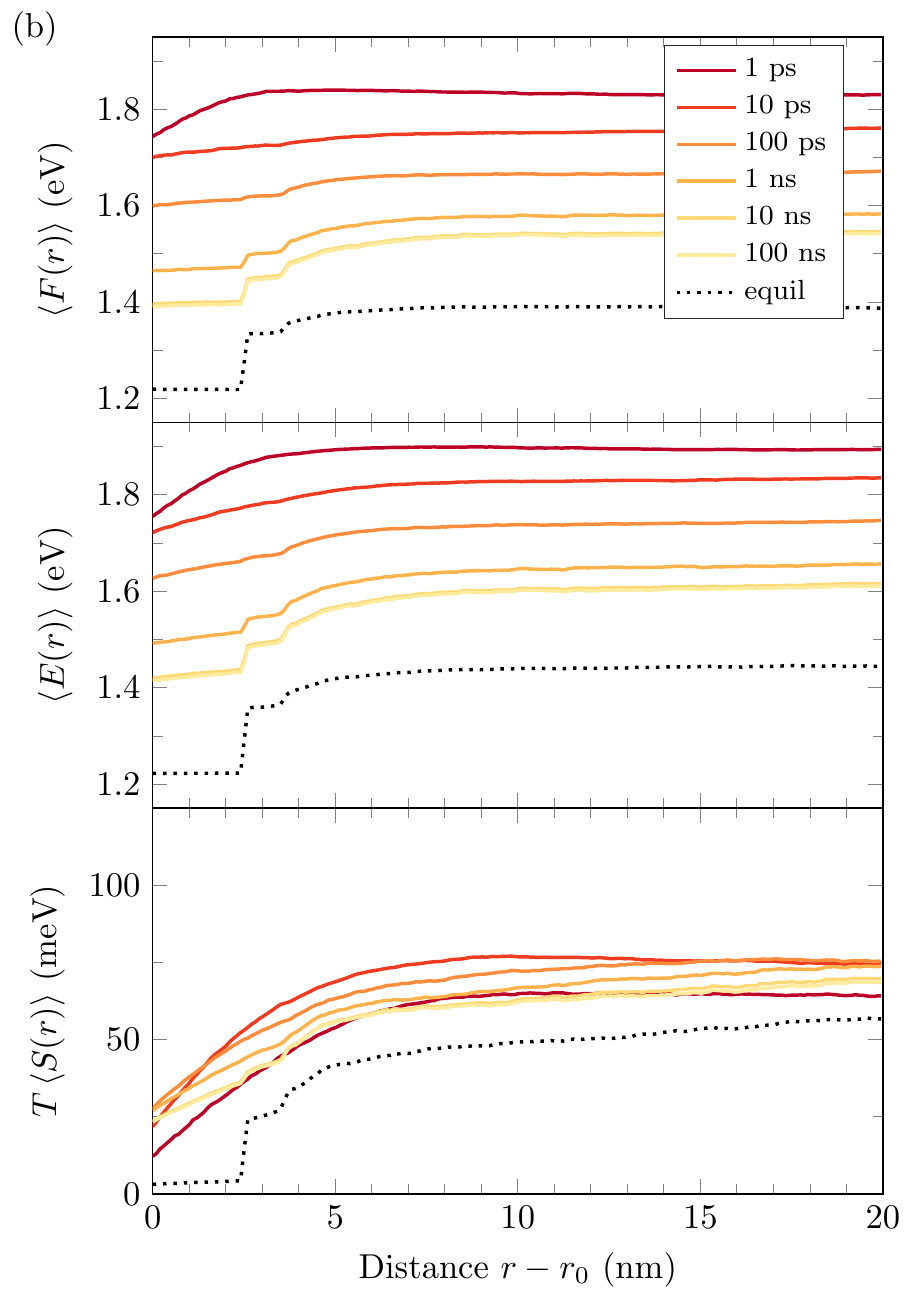}
    \caption{Time dependence of the free energy and its energy and entropy contribution for the parameter set (left) \textbf{all\_transfer\_less\_2} and (b)  \textbf{all\_transfer\_less\_4}.}
    \label{fig:time_dep_disorder}
\end{figure}

\pagebreak

\bibliography{references}